\documentclass[12pt]{article}
\begin{document}

\title{The Zitterbewegung Region}
\author{B.G. Sidharth\footnote{birlasc@gmail.com}, B.M. Birla Science Centre,\\ Adarsh Nagar, Hyderabad - 500 063, India\\
\\Abhishek Das\footnote{parbihtih3@gmail.com}, B.M. Birla Science Centre,\\ Adarsh Nagar, Hyderabad - 500 063, India\\}
\maketitle

\begin{abstract}
This paper deals with a precise description of the region of {\it zitterbewegung} below the Compton scale and the stochastic nature associated with it. We endeavour to delineate this particular region by dint of {\it Ito's calculus} and instigate certain features that are in sharp contrast with conventional physics.
\end{abstract}
\maketitle

\section{Introduction}

The phenomenon of {\it zitterbewegung} has been a subject of widespread interest since Schrodinger \cite{Erwin} proposed the said concept. Many authors \cite{huang,David,Barut} including the author Sidharth \cite{bgs1,bgs2} have studied such a phenomenon extensively. Particularly, the affinity of the {\it zitterbewegung} and the Compton scale has been studied by Sidharth, where the Compton scale has been considered to be elementary rather than the Planck scale.\\
The approach presented in this paper is distinguished in the sense that we identify the entire space to be comprised of an interior and an exterior region separated by a Jordan curve, which has been discussed in the second section. Our endeavours are related to the interior region which is must be treated in a meticulous and precise manner. The methodology used is that which was introduced by Kiyosi Ito \cite{Ito1}. Ito's calculus has been implemented to describe the stochastic nature of the interior region in the section 3. \\
Section 4 deals with the various transforms (Laplace, Fourier) that are duly modified in the interior region, on account of Ito's stochastic nature. The fifth section brings out the junction of classical and quantum physics that arises in this interior region and some other related aspects.\\
Finally, we discuss the various aspects of the interior region and how it can be looked upon as the region of juncture between classical and quantum mechanics. The quantum mechanical spin has been shown to arise from our considerations in the Compton scale, which therefore establishes the Compton scale in a fundamental level.

\section{The Jordan Curve}

Apart from the previous papers by Sidharth, in some other papers \cite{bgs3,bgs4,bgs5} it was found that there is indeed a transition from the {\it Planck scale} to the {\it Compton scale}, from whence we have the {\it coherence length} or {\it coherence parameter} \cite{bgs6,bgs7} as\\
\[\xi = \frac{h}{mc} = \frac{2\pi\hbar}{mc}\]\\
Or,\\
\[\xi = 2\pi l_{c}\]\\
where, $l_{c} = \frac{\hbar}{mc}$ is the Compton length of a particle. Now, this can be looked upon as the circumference of a circle of radius $l_{c}$ or merely, a nonintersecting closed curve whose length is $2\pi l_{c}$. In a generalized manner, we can say that this is a {\it Jordan curve} ($J$) that has two connected components: one being the {\it interior region} ($I$) bounded by $J$ and the other being the {\it exterior region} ($E$). \\
Now, since the sets $E$ and $I$ are disjoint and $J$ is the boundary separating them, we can infer that together they form a disconnected set $S$ which is the set of the whole space. Hereby, we begin our theory by making the presumption that the interior region ($I$) can be visualized as being below the Compton scale and the exterior region ($E$) being above the Compton scale, separated by the {\it Jordan curve} ($J$) whose length is given by the {\it coherence parameter} as above. This {\it coherence parameter} ($\xi$) being related to the Compton length is the measure of a {\it fundamental minimum length}.\\

Suppose, the state describing the exterior ($E$) is $\psi_{E}$ and that describing the interior ($I$) is $\psi_{I}$. Therefore, if $\psi_{S}$ is the state describing the entire space then we can write\\
\begin{equation}
\psi_{S} = \psi_{E} + \psi_{I}
\end{equation}
Now, if $\psi_{S1}$ and $\psi_{S2}$ be represent two different states of the set $S$ with probability $P_{S1}$ and $P_{S2}$. Thus, the density operator would be given as\\
\begin{equation}
\rho = P_{S1} |\psi_{S1}><\psi_{S1}| + P_{S2} |\psi_{S2}><\psi_{S2}|
\end{equation}
Using equation (1) we have\\
\begin{equation}
\rho = P_{S1} |\psi_{E1} + \psi_{I1}><\psi_{E1} + \psi_{I1}| + P_{S2} |\psi_{E2} + \psi_{I2}><\psi_{E2} + \psi_{I2}|
\end{equation}
Here, we would like to mention that according to the Feshbach-Villars approach \cite{Feshbach} Sidharth had shown \cite{bgs8,bgs9} that if we invoke a two component wavefunction\\
\[\psi = \left(\matrix {\phi\cr \chi}\right)\]\\
where $\phi$ and $\chi$ represent the positive and negative energy solutions respectively, then above the Compton scale the $\phi$ component is predominant and below the Compton scale the $\chi$ component is predominant. In our approach, we infer that the $\phi$ component dominates in the $E$-region and the $\chi$ component dominates in the $I$-region, although both components are present in the said regions.\\
Now as we can see from equation (3), the state describing the interior region ($I$) is present in the density matrix. So, if the space $S$ is studied on the whole then all we get is a mixed description of the two regions. Now, this is not feasible as we shall argue later, because the interior region is in it's entirety a completely different picture. Due to {\it zitterbewegung} effects the processes going on in the region $I$ is stochastic and in this sense it is distinguished from the region $E$. The {\it Jordan curve} ($J$) not only separates the $I$-region from the $E$-region, but also poses some restrictions on the region $I$. In a nutshell, this implies that the methodology used to describe the $E$-region cannot be implemented in case of the $I$-region. The {\it Jordan curve} ($J$) whose magnitude is related to the Compton length binds the stochastic effects below the Compton scale. Consequently, equation (2) will not be able to provide a feasible description of the state of the whole space $S$. Also, the processes going on in the $E$-region will not be affected by those of the $I$-region. \\
All of this suggests that the regions $E$ and $I$ must be studied separately in a meticulous manner, in order to get a better conception of the space $S$ on the whole. Again, it is interesting to point out the fact that if the space $S$ was connected, i.e. the curve $J$ didn't exist or more precisely the {\it coherence parameter} wasn't epitomized as a {\it fundamental minimum length} then we would simply have the density operator as\\
\begin{equation}
\rho = P_{S1} |\psi_{S1}><\psi_{S1}| + P_{S2} |\psi_{S2}><\psi_{S2}|
\end{equation}
where, $\psi$ simply equals to $\psi_{S}$. The entire space $S$ would be described merely by $\psi_{S}$. But, as we know matters aren't so simple and thus we would like to study elaborately the stochastic processes going on inside the region $I$.\\

\section{The Interior Region}

As we mentioned in the preceding section, the interior region ($I$) is a bounded region. All the processes going on in it are stochastic due to {\it zitterbewegung} effects. The processes could be classical or quantum in nature but they are always stochastic and epitomize the uniqueness of the $I$-region. Also, these can be looked upon as {\it Ito processes} ($p$) \cite{Ito1,Ito2} culminating from Brownian-type motion \cite{Wim} of the particles. Suppose, $\phi_{t}$ describes the processes $p_{t}$ at the instant '$t$' such that\\
\[\phi_{t} = \exp(\pm p_{t})\]\\
Or,
\begin{equation}
p(\phi_{t}) = p_{t} = \mp \ln\phi_{t}
\end{equation}
where, $\phi_{t}$ is itself a stochastic process due to random fluctuations going on in every aspect. Also, we consider a {\it Wiener process} $W_{t}$ \cite{Wim} which itself is a stochastic process such that\\
\begin{equation}
W_{t} = \int_{0}^{t} f(p_{t}){\rm d}\phi_{t}
\end{equation}
This is {\it Ito's integral} \cite{Ito1,Anders} which defines {\it Ito's calculus}  and $f(p_{t})$ is another process which is adapted to the filtration generated by $\phi_{t}$. Here 'filtration' means the set of information available up to the time '$t$'. This means that a stochastic process depends on the information available till the instant the said process occurs. More precisely, a stochastic process depends on the previous stochastic processes. We shall endeavour to prove this more rigorously.\\

Let us consider {\it Ito's lemma} for the stochastic {\it drift diffusion} processes where the process $p(\phi_{t})$ do not explicitly depend on time ($\frac{\partial p(\phi_{t})}{\partial t} = 0$), such that\\
\begin{equation}
{\rm d}p(\phi_{t}) = p^{\prime}(\phi_{t}){\rm d}\phi_{t} + \frac{1}{2} p^{\prime\prime}(\phi_{t})\sigma_{t}^{2}\phi_{t}^{2}{\rm d}t
\end{equation}
where, the primes ($\prime$) denote partial derivatives with respect to $\phi_{t}$ and the expectation \\
\[E[{\rm d}\phi_{t}^{2}] = \phi_{t}^{2}{\rm d}t\]\\
In parallel to this, as an analogue to methods used in stochastic mathematical finance \cite{Mall} we consider the processes ($\phi_{t}$) to be characterized by {\it geometric Brownian motion}, such that\\
\begin{equation}
{\rm d}\phi_{t} = \phi_{t}[\mu{\rm d}t + \sigma_{t}{\rm d}W_{t}]
\end{equation}
where, $\mu$ is the mean or {\it drift} and $\sigma_{t}$ is an infinitesimal standard deviation. Now, in the $E$-region where stochastic processes have no relevance ($W_{t} = 0$), this equation will be simply\\
\[{\rm d}\phi = \phi\mu{\rm d}t\]\\
and with ordinary calculus have the solution\\
\begin{equation}
\phi_{t} = \phi_{0}\exp[\pm\mu t]
\end{equation}
But, in the $I$-region when we resort to Ito's stochastic calculus the solution differs a lot. Putting equation (8) in (7) we have\\
\begin{equation}
{\rm d}p_{t} = p^{\prime}\phi_{t}[\mu{\rm d}t + \sigma_{t}{\rm d}W_{t}] + \frac{1}{2} p^{\prime\prime}\sigma_{t}^{2}\phi_{t}^{2}{\rm d}t
\end{equation}
Using relation (5) we have\\
\begin{equation}
\mp{\rm d}(\ln\phi_{t}) = \mu{\rm d}t + \sigma_{t}{\rm d}W_{t} \pm \frac{1}{2}\sigma_{t}^{2}{\rm d}t
\end{equation}
Thus, we obtain finally the solution\\
\begin{equation}
\phi_{t} = \phi_{0}\exp[\pm\{\sigma_{t}W_{t} + (\mu \pm \frac{\sigma_{t}^{2}}{2})t\}]
\end{equation}
It is obvious that this solution is entirely different from that of the solution give in equation (9). This is a stringent reason as to why the $I$-region must be looked upon differently from than the $E$-region. Due to the stochastic nature of the processes going on in the $I$-region the use of Ito's calculus yields different results \cite{Hui}.\\
Now, in terms of partial derivatives we can write equation (8) as\\
\[\frac{\partial W_{t}}{\partial \phi_{t}} = \frac{1}{\sigma_{t}}[\frac{1}{\phi_{t}} - \frac{\mu}{\frac{\partial \phi_{t}}{\partial t}}]\]\\
Again, using the solution (12) this gives us\\
\begin{equation}
\frac{\partial W_{t}}{\partial \phi_{t}} = \frac{1}{\sigma_{t}}[\frac{1}{\phi_{t}} - \frac{\mu}{\phi_{0}(\mu - \frac{\sigma_{t}^{2}}{2})}]
\end{equation}
This can also be written as\\
\[\frac{\partial W_{t}}{\partial \phi_{t}} = f(\phi_{t})\]\\
where, $f(\phi_{t}) = \frac{1}{\sigma_{t}}[\frac{1}{\phi_{t}} - \frac{\mu}{\phi_{0}(\mu - \frac{\sigma_{t}^{2}}{2})}]$. Therefore, we have\\
\begin{equation}
W_{t} = \int_{0}^{t} f(\phi_{t}) {\rm d}\phi_{t}
\end{equation}
where, $f(\phi_{t})$ is another stochastic Ito process. This methodology can be continued to find a succession of such stochastic processes. Also, the process $f(p_{t})$ in equation (6) can be traced back to another stochastic Ito process. Therefore, we can conclude that in the $I$-region there are plethora of stochastic processes going on and all of them are connected to each other. The chain of such processes carries on indefinitely as one process triggers the start of another process and so on. This is the underlying feature of the $I$-region that distinguishes it from the $E$-region and all of this is caused due to the {\it Jordan curve} ($J$) that acts as the barrier between the aforesaid regions.\\
Now, let us consider any two arbitrary processes $p(\phi_{t1})$ and $p(\phi_{t2})$ occurring at two instants $t_{1}$ and $t_{2}$. If $Pr[p(\phi_{t1})]$ and $Pr[p(\phi_{t2})]$ are the individual probabilities of the processes then in terms of conditional probability it is obvious from our results that\\
\begin{equation}
Pr[p(\phi_{t2})] \neq Pr[p(\phi_{t2})|p(\phi_{t1}))]
\end{equation}
since the processes are connected. Thus, we can conclude that all the processes in the $I$-region are dependent of one another. Again, the processes in the $E$-region are not connected to those in the $I$-region. Therefore, for a process $X$ in the $E$-region and a process $p(\phi_{t})$ in the $I$-region we would have\\
\begin{equation}
Pr[X] = Pr[X|p(\phi_{t})]
\end{equation}
Consequently, as we mentioned earlier, when one studies the space $S$ the regions $E$ and $I$ must be studied separately and with relevant methods: ordinary calculus for the $E$-region and Ito's calculus for the $I$-region. Otherwise, there would be discrepancies and one would not be able to get the true picture of the whole space. The reason is obviously the stochastic nature of the $I$-region which emanates from the different nature of the physics inside the $I$-region.\\

\section{The Laplace transform in the {\it I}-region}

Now, let us consider that the interior ($I$) region is a subset of {\bf C}, the complex plane. If '$s$' and '$t$' denote the frequency and time respectively then the Laplace transform of a function $f(t)$ is be given as\\
\begin{equation}
F(s) = L[f(t)] = \int_{0}^{\infty} e^{-st}f(t) {\rm d}t
\end{equation}
Now, suppose that in the $I$-region $f(t)$ and $F(s)$ represent stochastic processes as functions of time and frequency respectively. Then, we would like to see the nature of this transform when {\it Ito's calculus} is taken into consideration. One may write equation (17) as\\
\[{\rm d}F(s) = e^{-st}f(t){\rm d}t\]\\
But, for the $I$-region if we consider {\it Ito's calculus} then we may write\\
\begin{equation}
{\rm d}F(s) = \dot{F}(s){\rm d}t + \frac{1}{2}F^{\prime\prime}(s)\sigma_{t}^{2}s^{2}{\rm d}t
\end{equation}
where the prime ($\prime$) denotes derivative with respect to the frequency ($s$) and the {\it dot} denotes partial derivative with respect to time ($t$), assuming that the $F(s)$ depends explicitly on '$t$' only ($\frac{\partial F(s)}{\partial s} = 0$). Here\\
\[\dot{F}(s) = e^{-st}f(t)\] \\

Now, in the region outside $I$, the second term on the right hand side of equation (18) does not exist because there we have the conventional calculus. But, in the $I$-region considering {\it Ito's calculus} the second term is of non-trivial importance. Integrating equation (18) we have\\
\[F(s) = \int_{0}^{\infty} F^{\prime}(s){\rm d}t + \frac{\sigma_{t}^{2}}{2}\int_{0}^{\infty} s^{2}\frac{\rm d}{{\rm d}s}[e^{-st}f(t)]{\rm d}t\]\\
Assuming, the frequency is independent of the integration with respect to time, this gives\\
\[F(s) = L[f(t)] - \frac{\sigma_{t}^{2}}{2}s^{3} \int_{0}^{\infty} e^{-st}f(t) {\rm d}t + \frac{\sigma_{t}^{2}}{2}s^{2} \int_{0}^{\infty} e^{-st}h(t) {\rm d}t\]\\
where, $h(t) = f^{\prime}(t)$ is another stochastic processes. Therefore, we have \\
\begin{equation}
F(s) = L[f(t)] + \frac{\sigma_{t}^{2}}{2} (s^{2}L[h(t)] - s^{3}L[f(t)])
\end{equation}
More rigorously, this can be written as\\
\begin{equation}
F(s) = L[f(t)] + \frac{\sigma_{t}^{2}}{2} G(s, t)
\end{equation}
where, $G(s, t) = s^{2}L[h(t)] - s^{3}L[f(t)]$ is another stochastic process. This result substantiates that in the $I$-region the expression for the Laplace transform is modified due to the host of stochastic processes going on. Ostensibly, outside the $I$-region we would have $\sigma_{t} = 0$, and consequently then we would get back the original Laplace transform. Now, from equation (19) we can also infer that the Laplace transform of a stochastic process in the complex plane is dependent of other stochastic processes going on in the $I$-region. This is exactly in favour of what we derived in the previous section.\\
Another interesting conclusion that we can draw from this result is that for other transforms also such stochastic effects would be present. Since, the Fourier, Mellin transforms are related to the bilateral or two-sided Laplace transform one can easily conclude a stochastic nature of the Fourier and Mellin transforms. The same can be deduced for the Borel transform and the z-transform.\\

\section{The juncture of Classical and Quantum mechanics}

Although, the present approach is purely classical it is interesting to point out the fact that the author Sidharth had used a slightly similar technique distinguishing the interior region from the exterior region. Averaging over the interior region one arrives at the domain of conventional physics. In fact, the author was also able to derive the Compton length from the classical point of view, averaging over the physically inaccessible interior region. This coincides with our approach, since we have considered the magnitude of the boundary curve $J$ to be related to the Compton length itself, via the {\it coherence parameter} ($\xi$).\\
In accordance with a previous work of the author Sidharth \cite{bgs10} and Moller \cite{Moller} we would like to stress another key feature of the interior region. It can be shown that in the relativistic case, if we consider the $I$-region as the rest system then we have a host of mass centres ($C_{i}$'s) that form a two dimensional circular disk perpendicular to the direction of the angular momentum ($L = l \times p$) of the system with centre at the proper centre of mass. The radius of such a disk is given as\\
\[r = \frac{L}{mc} = \frac{l \times p}{mc}\]\\
Now, if the momentum corresponds to the De Broglie wavelength of the system ($I$-region) such that\\
\[p = \frac{h}{l}\]\\
then, we have\\
\begin{equation}
r = 2\pi \frac{\hbar}{mc} = 2\pi l_{c}
\end{equation}
which is exactly the {\it coherence parameter} ($\xi$). Since, the Compton length ($l_{c}$) is a purely quantum mechanical we find that due to relativistic considerations classical and quantum physics have an affine connection. More interestingly, it was also shown by the author Sidharth \cite{bgs10,bgs11} that we can indeed obtain the quantum mechanical spin starting from classical physics, in the Compton region which is basically the $I$-region in this paper. As stated by Sidharth in light of Dirac's approach \cite{Dirac}, the interior region represents itself as a doubly connected space which has a nodal singularity or quantized vortex which gives rise to the quantum mechanical spin. The $I$-region is inherently the region of {\it zitterbewegung} and inside this region we realize the juncture of classical and quantum mechanics.\\
We would like to obtain further insight regarding this subject. Now, it is known that for two different solutions $\psi_{1}(x)$ and $\psi_{2}(x)$ of the Dirac equation\\
\[(\gamma^{\mu}p^{\mu} - mc)\psi(x) = 0\]\\
we have due to linearity \cite{Bjorken}\\
\begin{equation}
c\bar{\psi}_{2}\gamma^{\mu}\psi_{1} = \frac{1}{2m}[\bar{\psi}_{2}p^{\mu}\psi_{1} - (p^{\mu}\bar{\psi}_{2})\psi_{1}] - \frac{i}{2m}p_{\nu}(\bar{\psi}_{2}\sigma^{\mu\nu}\psi_{1})
\end{equation}
where, $\gamma^{\mu}$'s are the gamma matrices and $\sigma^{\mu\nu} = \frac{i}{2}[\gamma^{\mu}, \gamma^{\nu}]$. It is also known that this equation leads to the considerations of both positive and negative energy solutions. Now, in equation (22) we identify the term consisting $\sigma^{\mu\nu}$ as the term arising due to the motion about the proper centre of mass (as discussed above). Consequently, the momentum associated with it can be identified as the spin angular momentum due to the presence of $\sigma^{\mu\nu}$ which consists of the gamma matrices. Interestingly, calculating the average current which is given by the expectation value of the velocity operator $c\alpha$ [where, $\alpha_{i} = \left(\matrix {0 & \sigma_{i}\cr \sigma_{i} & 0}\right)$] yields cross terms between both the positive and negative energy solutions that fluctuate rapidly leading to the phenomenon of {\it zitterbewegung}. \\
These negative energy solutions are localized within the $I$-region (of radius or extension $l_{c}$) and bounded by the {\it Jordan curve} (of length = $2\pi l_{c}$). Therefore, we see that the quantum mechanical spin arises quite elegantly from our considerations in the $I$-region, when we take into account the motion about the centre of proper mass of the system. Now, it must be stringently borne in mind that outside the $I$-region the negative energy solutions are dominated by the positive energy solutions and they are no longer localized. Consequently, the phenomenon of {\it zitterbewegung} has no effects in the $E$-region.

As we discussed previously, it was shown by the author Sidharth \cite{bgs8,bgs9} that the in the $I$-region the negative energy solutions dominate whereas in case of the $E$-region the positive energy solutions dominate. This feature of the $I$-region was also discussed recently \cite{bgs7} from the context of an {\it Ising-like model}.

\section{Discussions}
1)~Continuing in the vein of Section 5, we observe that as John Wheeler
put it, it is the spin half that separates Quantum Mechanics from
Classical Physics \cite{wheeler}. The spin half is also connected
with the doubly connected space feature of Quantum Theory as brought
out by Fermionic four spinor wave functions. More explicitly
\cite{schweber}, a scalar is a one component object of the rotation
rule, under which
$$\xi \to \xi' = \xi$$
On the other hand a spinor (of rank 1) is a two component object
which transforms according to
\begin{equation}
\xi \to \xi' = (1 + \frac{1}{2} \imath \epsilon \sigma_{i})
\xi\label{110}
\end{equation}

under the rotation rule, i.e. a rank 1 spinor gets transformed by
a $2 \times 2$ unitary matrix (determinant = 1). In the same vein a
vector is a three component object. In the case of the Quantum Mechanical four spinor,
$$\left(\begin{array}{ll}
\phi\\
\chi\end{array}\right)$$ alluded to above, the two spinor $\chi$
transforms according to
$$\chi \to - \chi$$
under reflections while
$$\phi \to \phi$$
under the same reflection. Here, four dimensional reflection of the Dirac
matrices is essential because the usual laws of electromagnetic
theory are invariant under reflections \cite{heine}.\\
The Quantum Mechanical wave function defines a doubly connected
space. In this sense, rather than ordinary physics simply connected
with space, herein comes the difference between Classical Physics
and Quantum Physics.\\
2)~Also, an interesting fact can be mentioned in accordance with our work. The author Sidharth in some of his papers \cite{bgs8,bgs9} had considered a double Wiener process (random forward-backward motion of time) that lead to a complex velocity $V - \imath U$, which means that the conventional space coordinate $x$ is replaced as\\
\[x \rightarrow x + \imath x^{\prime}\]\\

where, $x^{\prime}$ is an arbitrary function of time, i.e. we have a new imaginary coordinate (in 1D). Let us discuss this a little elaborately. It has been show by the author Sidharth that within the Compton scale, the {\it zitterbewegung} or self interaction effects gives rise to the inertial mass. In such a fundamental domain the fractal nature of space-time and a stochastic underpinning are found to be correlated. Particularly, in our work the $I$-region is such a domain and there we have time as a back and forth motion leading to a double Wiener process. In the vein of such stochastic nature of time the forward and backward velocities can be defined as \\
\[\frac{{\rm d}_{+}}{{\rm d}t}x(t) = b_{+}\]\\
\[\frac{{\rm d}_{-}}{{\rm d}t}x(t) = b_{-}\]\\

This leads to the well known Fokker-Planck equations [] as below\\
\begin{equation}
\frac{\partial \rho}{\partial t} + \nabla(\rho b_{+}) = V\Delta\rho
\end{equation}
and\\
\begin{equation}
\frac{\partial \rho}{\partial t} + \nabla(\rho b_{-}) = -U\Delta\rho
\end{equation}
where, we have defined\\
\[V = \frac{b_{+} + b_{-}}{2}; U = \frac{b_{+} - b_{-}}{2}\]\\
Using equations (1) and (2) we have\\
\begin{equation}
\frac{\partial \rho}{\partial t} + \nabla(\rho V) = 0
\end{equation}
and
\begin{equation}
U = \nu\nabla\ln\rho
\end{equation}
where, $\nu = lv$ is the diffusion constant, $l$ is the mean free path and $v$ is the velocity. Here, $V$ and $U$ are the statistical averages of the corresponding velocities. Introducing the following identifications\\
\begin{equation}
V = 2\nu\nabla S
\end{equation}
and
\begin{equation}
V - \imath U = -2\imath\nu\nabla(\ln \psi)
\end{equation}
we get the complex wavefunction
\begin{equation}
\psi = \sqrt{\rho}e^{\frac{\imath S}{\hbar}}
\end{equation}
which is the solution to the Schrodinger wave equation of quantum mechanics. Now, as can be seen we have introduced the complex velocity  $V -\imath U$ which insinuates that the ordinary coordinates will get complexified as below:\\
\[x \rightarrow x + \imath x^{\prime}\]\\
where, the real part of the velocity would be\\
\[\frac{{\rm d}X_{r}}{{\rm d}t} = V\]\\
and the imaginary part would be given as\\
\[\frac{{\rm d}X_{i}}{{\rm d}t} = U\]\\
Therefore, we have the total Wiener-type velocity as\\
\begin{equation}
W = \frac{{\rm d}}{{\rm d}t}(X_{r} - \imath X_{i})
\end{equation}
Taking, $x^{\prime} = ct$ we finally find that the Minkowski coordinate system ($x, \imath ct$) emerges very elegantly and so does special relativity. One may apply the methodology of {\it quaternions} to generalize this concept to 3D. As it was shown by the author Sidharth that this leads to four dimensions automatically, considering the Pauli matrices. \\
Thus, due to the random nature of a double Wiener process, the coordinates get complexified inside the $I$-region and as a consequence we obtain the roots of inertial mass, special relativity and quantum mechanics. This alone emphasizes the fundamental nature of the $I$-region and the Jordan curve (boundary of a fundamental minimum length).\\
It is worth noting that equation (7) resembles a covariant
derivative that epitomizes the geometry of the $I$-region itself.
From this covariant derivative one can understand the underlying
fundamental structure of space-time.\\
Also, many years ago it was shown by Ezra Newman \cite{Ezra} that if one considers the extension of Maxwell's equations into a complex Minkowski space and their invariance under the complex Poincare group then new solutions can be found. The precise value of the gyromagnetic ratio ($g = 2$) according to Dirac's theory can also be explained elegantly. Actually, from the derivations of our current paper we can indeed infer that such fundamental features of quantum mechanics begin in the $I$-region and are extended to the $E$-region, but with stochastic effects distinguishing the two different regions of physics.

3)~To see all this in the context of Dirac's nodal singularity let us
start with
\begin{equation}
\psi = \psi_1 e^{\imath S},\label{5xe2}
\end{equation}
where $\psi_1$ is the usual wave function with what may be called an
integrable phase. Further the phase $S$ does not have a definite
value at each point. The four gradient
\begin{equation}
K^\mu = \partial^\mu S\label{5xe3}
\end{equation}
is well defined. We use temporarily natural units, $\hbar = c = 1$.
Dirac then goes on to identify $K$ in (\ref{5xe3}) (except for the
numerical factor $hc/e$) with the electromagnetic
field potential, as in the Weyl gauge invariant theory.\\
Next Dirac considered the case of a nodal singularity, which is
closely related to what was later called a quantized vortex a term
we have already noted (Cf. for example ref.\cite{vasu}). In this
case a circuit integral of a vector as in (\ref{5xe3}) gives, in
addition to the electromagnetic term, a term like $2 \pi n$, so that
we have for a change in phase for a small closed curve around this
nodal singularity,
\begin{equation}
2 \pi n + e \int \vec B \cdot d \vec S\label{5xe4}
\end{equation}
In (\ref{5xe4}) $\vec B$ is the magnetic flux across a surface
element $d \vec S$ and $n$ is the number of nodes within the
circuit. The expression (\ref{5xe4}) directly lead to the
Monopole in Dirac's formulation.\\
Let us now reconsider the above arguments in terms of our earlier
developments. As we saw the Dirac equation for a spin half particle
throws up a complex or non Hermitian position coordinate
\cite{bgsfpl172004,bgsfpl162003}. Dirac as noted identified the
imaginary part with zitterbewegung effects and argued that this
would be eliminated when averages over intervals of the order of the
Compton scale are taken to recover meaningful physics
\cite{diracpqm}. Over the decades the significance of such cut off
space time intervals has been stressed by T.D. Lee and several other
scholars \cite{bgs10,kard,bom,leepl}.
Indeed we saw that with a minimum cut off length $l$, it was shown
by Snyder that there would be a non commutative but Lorentz
invariant spacetime structure. At the Compton scale we would have
\cite{uof},
\begin{equation}
[x,y] = 0(l^2)\label{5xe5}
\end{equation}
and similar relations.\\
In fact starting from the Dirac equation itself, we deduced directly
the
non commutativity (\ref{5xe5}) (Cf.refs.\cite{bgsfpl172004,bgsfpl162003}).\\
Let us now return to Dirac's formulation of the monopole in the
light of the above comments. As noted above, the non integrability
of the phase $S$ in (\ref{5xe2}) gives rise to the electromagnetic
field, while the nodal singularity gives rise to a term which is an
integral multiple of $2 \pi$. As is well known \cite{heap} we have
\begin{equation}
\vec \nabla S = \vec p\label{5xe6}
\end{equation}
where $\vec p$ is the momentum vector. When there is a nodal
singularity, as noted above, the integral over a closed circuit of
$\vec p$ does not vanish. In fact in this case we have a circulation
given by
\begin{equation}
\Gamma = \oint \vec \nabla S \cdot d \vec r = \hbar \oint dS = 2 \pi
n\label{5xe7}
\end{equation}
It is because of the nodal singularity that though the $\vec p$
field is irrotational, there is a vortex - the singularity at the
central point associated with the vortex makes the region multiply
connected, or alternatively, in this region we cannot shrink a
closed smooth curve about the point to that point. In fact if we use
the fact as seen above that the Compton wavelength is a minimum cut
off, then we get from (\ref{5xe7}) using (\ref{5xe6}), and on taking
$n = 1$,
\begin{equation}
\oint \vec \nabla S \cdot d\vec r = \int \vec p \cdot d \vec r =
2\pi mc \frac{1}{2mc} = \frac{h}{2}\label{5xe8}
\end{equation}
$l = \frac{\hbar}{2mc}$ is the radius of the circuit and $\hbar = 1$
in the above in natural units. In other words the nodal singularity
or quantized vortex gives us the mysterious Quantum Mechanical spin
half (and other higher spins for other values of $n$). In the case
of the Quantum Mechanical spin, there are $2 \times n/2 + 1 = n + 1$
multiply connected regions, exactly as in the case of nodal
singularities.

\end{document}